\documentclass[11pt]{article}
\usepackage{amsmath,amssymb}
\usepackage{graphicx}
 \allowdisplaybreaks
 \newcommand{\be}{\begin{equation}}
 \newcommand{\ee}{\end{equation}}
 \newcommand{\bea}{\begin{eqnarray}}
 \newcommand{\eea}{\end{eqnarray}}
 \newcommand{\ds}{\displaystyle}
 \newcommand{\nn}{\nonumber}
 \newcommand{\td}{\tilde}
 \newcommand{\wtd}{\widetilde}
 \newcommand{\pd}{\partial}
 \newcommand{\one}{{\bf 1}}
 
 \newcommand{\bm}{{\bf m}}
 \newcommand{\bn}{{\bf n}}
 
 \newcommand{\bk}{{\bf k}}

 \newcommand{\bw}{{\bf w}}
 
 \newcommand{\bJ}{{\bf J}}
 \newcommand{\bL}{{\bf L}}
 
 \newcommand{\bQ}{{\bf Q}}
 \newcommand{\cA}{{\cal A}}

 \newcommand{\cL}{{\cal L}}
 \newcommand{\cM}{{\cal M}}
 
 \newcommand{\cO}{{\cal O}}

 \newcommand{\cV}{{\cal V}}
 
 \newcommand{\tda}{{\td a}}
 \newcommand{\tdb}{{\td b}}

 \newcommand{\lie}{\pounds}
\long\def\symbolfootnote[#1]#2{\begingroup%
\def\thefootnote{\fnsymbol{footnote}}\footnote[#1]{#2}\endgroup}
\newcommand{\aei}{\it Max Planck Institute for Gravitational Physics
(Albert Einstein Institute)\\ Am M\"uhlenberg 1, D-14476 Golm, Germany}

\newcommand{\auth}{Jianwei Mei}

\begin{document}
%%%%%%%%%%%%%%%%%%%%%%%%%%%%%%%%%%%%%%%%%%%
\thispagestyle{empty}
%%%%%%%%%%%%%%%%%%%%%%%%%%%%%%%%%%%%%%%%%%%
\begin{flushright}
\hfill{AEI-2012-015}
\end{flushright}
%%%%%%%%%%%%%%%%%%%%%%%%%%%%%%%%%%%%%%%%%%%
\begin{center}

~\vspace{20pt}

{\Large\bf Conformal Symmetries of the Einstein-Hilbert Action on
Horizons of Stationary and Axisymmetric Black Holes}

\vspace{25pt}

\auth \symbolfootnote[1]{Email:~\sf jwmei@aei.mpg.de}

%\vspace{10pt}{\ihep}

%\vspace{10pt}{\tamu}

\vspace{10pt}{\aei}

\vspace{2cm}

\underline{ABSTRACT}

\end{center}

We suggest a way to study possible conformal symmetries on black
hole horizons. We do this by carrying out a Kaluza-Klein like
reduction of the Einstein-Hilbert action along the ignorable
coordinates of stationary and axisymmetric black holes. Rigid
diffeomorphism invariance of the $m$-ignorable coordinates then
becomes a global $SL(m,R)$ gauge symmetry of the reduced action.
Related to each non-vanishing angular velocity there is a
particular $SL(2,R)$ subgroup, which can be extended to the Witt
algebra on the black hole horizons. The classical Einstein-Hilbert
action thus has $k$-copies of infinite dimensional conformal
symmetries on a given black hole horizon, with $k$ being the
number of non-vanishing angular velocities of the black hole.

%%%%%%%%%%%%%%%%%%%%%%%%%%%%%%%
 \newpage
% \setcounter{footnote}{0}
% \setcounter{page}{1}
%%%%%%%%%%%%%%%%%%%%%%%%%%%%%%%

\tableofcontents
%%%%%%%%%%%%%%%%%%%%%%%%%%%%%%%

\section{Introduction}

It is a long standing problem to find a statistical explanation of
the black hole entropy. One intriguing possibility is that the
black hole entropy may have a sort of ``universal" explanation,
which is largely determined by some 2D conformal filed theory but
depends little on the detail of the possible UV completion of
quantum gravity. Discussions of such an idea can be found in, e.g.
\cite{carlip07,bkls11}.

There have been some evidence in support of this possibility. Soon
after the original calculation of the entropy for certain black
holes in string theory \cite{strominger.vafa96}, Strominger showed
that any black holes having an $AdS_3$ factor in their near
horizon region can have their entropies calculated in a common way
\cite{strominger97}, by using the fact that quantum gravity on
$AdS_3$ must be described by a 2D conformal field theory (CFT)
\cite{bh86}. Loosely related to this, it has also been suggested
that, with appropriate boundary conditions imposed, quantum
gravity on the horizon of black holes may also be described by a
2D conformal field theory
\cite{carlip98,solodukhin98,carlip99,carlip11a,carlip11b}. This
later argument, however, is marred by the ambiguity on the
possible boundary conditions that one can impose near the black
hole horizons.

More recently, the development of the Kerr/CFT correspondence
\cite{ghss08,bkls11} brings more support to a possible
``universal" explanation of the black hole entropy. The near
horizon limit of the extremal Kerr (NHEK) metric
\cite{bardeen.horowitz99} at fixed polar angles are quotients of
warped $AdS_3$. This indicates that one may use the same
techniques of \cite{bh86} to discuss the asymptotic symmetry
group, much like in the case of BTZ black holes
\cite{strominger97}, which are quotients of $AdS_3$. Indeed, for
an extremal Kerr black hole with the angular momentum $J$,
appropriate boundary conditions can be found and a copy of the
Virasoro algebra can be identified. The putative CFT at the NHEK
boundary was shown to have a central charge $c_L=12J$ and
temperature $T_L=\frac1{2\pi}$ \cite{ghss08}. Cardy's formula then
reproduces exactly the Bekenstein-Hawking entropy. Afterwards, the
calculation was generalized to black holes in higher dimensions
and also in more complicated settings (for a sample of the early
references, see \cite{lu.mei.pope08, hmns08,
azeyanagi.ogawa.terashima08, chow.cvetic.lu.pope08,
lu.mei.pope.justin09, compere.murata.nishioka09, mty09}).  Black
holes in more than four dimensions can have multiple rotations. It
was found in \cite{chow.cvetic.lu.pope08} that corresponding to
each non-zero rotation there is an independent copy of the
Virasoro algebra, and each copy of the Virasoro algebra appears to
be equally good in reproducing the Bekenstein-Hawking entropy. For
general treatments, it has also been shown that the method works
for all extremal stationary and axisymmetric black holes, in the
context of Einstein gravity \cite{mei10}.

As a drawback, the success of the Kerr/CFT correspondence is
limited to extremal black holes. Although it is possible to
discuss physics slightly away from the extremal limit (see, e.g.
\cite{castro.larsen09}), it will be more desirable to study the
case of non-extremal black holes directly. The investigation of
the hidden conformal symmetry of Kerr black hole is one such
attempt \cite{cms10}. In steady of looking at the symmetry
structure of gravitational fluctuations directly, the authors of
\cite{cms10} studied the dynamics of a massless scalar field
probing the background of a Kerr black hole. They found that the
wave equation in the so called ``near region" enjoys an enhanced
$SL(2,R)_L\times SL(2,R)_R$ symmetry. By assuming that there is a
putative dual 2D CFT having a ground state sharing this same
$SL(2,R)_L\times SL(2,R)_R$ symmetry, the authors of \cite{cms10}
were able to infer for the temperatures $T_{L,R}$, which together
with the central charges $c_{L,R}$ extrapolated from the Kerr/CFT
calculation, reproduce the Bekenstein-Hawking entropy exactly.
Further evidence of the existence of a dual 2D CFT was also
provided by matching the low-energy scalar-Kerr scattering
amplitude with correlators of a 2D CFT at the same temperatures.
For further works one can consult \cite{cvetic.larsen11} and
references therein.

Still, the situation with non-extremal black holes is far from
being satisfactory. In order to achieve the same level of success
as is in the case of Kerr/CFT correspondence for extremal black
holes, one will need a way to identify the full conformal
symmetries of the putative dual 2D conformal field theory. In this
paper, we want to report some partial results that may finally
help us achieve this goal.

We will show that on the horizon of a stationary and axisymmetric
black hole with $k$ non-vanishing angular velocities, the
Einstein-Hilbert action itself enjoys $k$-copies of infinite
dimensional conformal symmetries. Note the similarity between this
result and that from \cite{chow.cvetic.lu.pope08} mentioned above.
Our result holds for any stationary and axisymmetric black holes
in any spacetime dimensions. But since we will limit our
calculation to pure Einstein gravity plus a (possibly zero)
cosmological constant, the black holes should also be solutions to
such a system.

Our starting point is the simple fact that stationary and
axisymmetric black holes all have ignorable coordinates and that
their metrics share a common structure \cite{mei10}. It is then
natural to seek a Kaluza-Klein like reduction of the action on the
ignorable coordinates. The usual experience with Kaluza-Klein
reduction suggests that it may be easier to study some of the
symmetries in the system (see, e.g. \cite{duff.nilsson.pope86,
julia.nicolai96}). On the other hand, since we presume the
existence of the classical black hole solutions, what we do here
is not much than explicitly writing out the classical action in
terms of functions that are known to be independent on the
ignorable coordinates. As such, we will not expect any
inconsistency that may arise in the usual Kaluza-Klein reduction
of a dynamical system. Rather, the reduced action allows us to
study the classical equations of motion in a much greater detail.
In the case of pure gravity plus a cosmological constant, this
allows us to re-derive the first law of black hole thermodynamics
in a straightforward manor. In fact, the derivation echoes with
\cite{gibbons.perry.pope04} and partially explains why it is
sensible to calculate the mass of a black hole by integrating the
first law of thermodynamics.

After the reduction, we find that the rigid diffeomorphism
invariance of the ignorable coordinates become a global $SL(m,R)$
gauge symmetry of the reduced action, with $m$ being the number of
the ignorable coordinates. As the key result of this paper, we
will show that corresponding to each non-vanishing angular
momentum there is a particular $SL(2,R)$ subgroup, which can be
extended to the full Witt algebra on the black hole horizons. This
means that the classical Einstein-Hilbert action, when restricted
to the horizons of stationary and axisymmetric black holes, enjoys
a copy of the infinite dimensional conformal symmetry for each
non-vanishing angular velocity.

The plan of the paper is as follows. In section 2, we derive a
scheme of Kaluza-Klein like reduction that will make it easier to
deal with the special case of stationary and axisymmetric black
holes. In section 3, we write down the reduced action for
stationary and axisymmetric black holes. As an application, we
re-derive the first law for black holes in terms of the new
language. In section 4, we prove the classical conformal
invariance of the reduced action on the black hole horizons. A
short summary is in section 5.

\section{A Kaluza-Klein Reduction of the Einstein-Hilbert Action}

Consider the action in an $D$-dimensional spacetime $\Sigma$ with
a boundary $\pd\Sigma$ ,
%%%
\be S=\int_\Sigma d^Dx\sqrt{|g|}\;(R-2\Lambda) +\int_{\pd\Sigma}
(d^{D-1}x)_\mu n^\mu\sqrt{|g|}\;K\,,\label{action}\ee
%%%
where $n^\mu$ is the unit normal vector of $\pd\Sigma$ (suppose
the boundary is defined with some function $\Delta=0$, then $n_\mu
=\pd_\mu\Delta/\sqrt{g^{\varrho\sigma} \pd_\varrho\Delta\pd_\sigma \Delta}$ ),
and $K$ is the extrinsic curvature,
%%%
\be K=g^{\mu\nu}K_{\mu\nu}\,,\quad K_{\mu\nu}=\nabla_\mu
n_\nu+\nabla_\nu n_\mu\,.\ee
%%%
The inclusion of the Gibbons-Hawking-York boundary term is
necessary for a well defined variation principle.
When the metric is varied (note $\delta g^{\mu\nu} =0$ on
$\pd\Sigma$),
%%%
\be\delta S=\int_\Sigma d^Dx \sqrt{|g|}\;\Big(R_{\mu\nu}
-\frac{R-2\Lambda}2 g_{\mu\nu} \Big)\delta g^{\mu\nu}\,,\ee
%%%
from which one can derive the equations of motion
%%%
\be R_{\mu\nu}=\frac{2\Lambda}{D-2}g_{\mu\nu}\,.\ee
%%%

Now consider the metric of a $(D=m+n)$-dimensional
spacetime,\footnote{Do not confuse the number $n$ with the normal
vector $n^\mu$ of the boundary $\pd\Sigma$.}
%%%
\be ds^2=\wtd{G}_{\mu\nu}dx^\mu dx^\nu=H_{IJ}dx^Idx^J
+G_{AB}dy^Ady^B\,,\label{metric}\ee
%%%
where both $H_{IJ}$ and $G_{AB}$ depend only on the
$x$-coordinates. We use capital letters from the beginning of the
alphabet ($A,B,C,\cdots\in\{1,\cdots,m\}$) to label the
$y$-coordinates, and those from the middle of the alphabet
($I,J,K,\cdots\in\{1,\cdots,n\}$) to label the $x$-coordinates.
The reason for considering such a metric will become clear in the
next section. Now because both $G_{AB}$ and $H_{IJ}$ depend only
on the $x$-coordinates, one can formally treat $G_{AB}$ as some
matter fields living in the curved background $H_{IJ}$. It is then
interesting to look at the action for both $G_{AB}$ and $H_{IJ}$
from this new perspective. For this purpose, let's write down the
metric elements explicitly,
%%%
\bea\wtd{G}_{IJ}=H_{IJ}\,,\quad \wtd{G}_{AB}
=G_{AB}\,,\quad\wtd{G}_{IA}=0\,,\nn\\
\Longrightarrow\quad \wtd{G}^{IJ}=H^{IJ}\,,\quad
\wtd{G}^{AB}=G^{AB}\,,\quad \wtd{G}^{IA}=0\,.\eea
%%%
From now on, indices $A,B,C,\cdots$ will be raised or lowered
using the metric $G$, and indices $I,J,K,\cdots$ will be raised or
lowered using the metric $H$. We will always write out the indices
$A,B,C,\cdots$ explicitly, but will sometimes hide the
$I,J,K,\cdots$ indices, in places where their presence is obvious.
The elements of the original affine connection are
%%%
\bea\wtd{\Gamma}^I_{JK}&=&\Gamma^I_{JK}\,,\quad
\wtd{\Gamma}^A_{IJ}=\wtd{\Gamma}^I_{AJ}
=\wtd{\Gamma}^A_{BC}=0\,,\nn\\
\wtd{\Gamma}^I_{AB}&=&-\frac12\pd^IG_{AB}\,,\quad
\wtd{\Gamma}^A_{IB}=\frac12G^{AC}\pd_IG_{BC}\,,\eea
%%%
the elements of the original Ricci tensor are
%%%
\bea\wtd{R}_{IJ}&=&R_{IJ}-\nabla_I\nabla_J\ln\sqrt{|G|}+\frac14
\pd_IG_{AB}\pd_JG^{AB}\,,\quad \wtd{R}_{IA}=0\,,\nn\\
\wtd{R}_{AB}&=&-\frac12\nabla^2G_{AB}-\frac12\pd \ln\sqrt{|G|}
~\pd G_{AB}+\frac12G^{CD}\pd G_{AC}\pd G_{BD}\,,
\label{elements.Ricci}\eea
%%%
and the original Ricci scalar is
%%%
\bea\wtd{R}&=&R-(\pd\ln\sqrt{|G|})^2-2\nabla^2
\ln\sqrt{|G|}+\frac14\pd G_{AB}\pd G^{AB}\nn\\
&=&R+(\pd\ln\sqrt{|G|})^2+\frac14\pd G_{AB}\pd G^{AB}
-\frac2{\sqrt{|G|}}\nabla^2\sqrt{|G|}\,.\eea
%%%
We will only consider the case when the boundary $\pd\Sigma$ is
in the $x$-directions. Then $\wtd{n}_A=0$, $\wtd{n}_I=n_I$
and
%%%
\bea\wtd{K}_{IJ}=K_{IJ}=\nabla_I\;n_J+\nabla_J\;n_I\,,\quad
\wtd{K}_{AB}=-2\wtd\Gamma^I_{AB}n_I=n_I\pd^IG_{AB}\,,\nn\\
\Longrightarrow\quad\wtd{K}=\wtd{H}^{IJ}\wtd{K}_{IJ}
+\wtd{G}^{AB}\wtd{K}_{AB}=K+2n^I\pd_I\ln\sqrt{|G|}\,.\eea
%%%
Using these results in the original action (\ref{action}), we find
%%%
\bea S&=&\int_\Sigma d^nx\sqrt{|H|}\sqrt{|G|}\; \Big\{R-2
\Lambda+(\pd\ln\sqrt{|G|})^2+\frac14\pd G_{AB}\pd G^{AB}
-\frac2{\sqrt{|G|}}\nabla^2\sqrt{|G|}\Big\}\nn\\
&&+\int_{\pd\Sigma}(d^{D-1}x)_I \;n^I\sqrt{|H|}
\sqrt{|G|}\;\Big\{K+2n^J\pd_J\ln\sqrt{|G|}\Big\}\,,\nn\\
&=&\int_\Sigma d^nx \sqrt{|H|}\sqrt{|G|}\;\Big\{R-2\Lambda+
(\pd\ln\sqrt{|G|})^2+\frac14\pd G_{AB}\pd G^{AB}\Big\}\nn\\
&&+\int_{\pd\Sigma}(d^{D-1}x)_I \;n^I\sqrt{|H|}
\sqrt{|G|}\;K\,,\label{action.reduced}\eea
%%%
where we have divided out the volume of the $y$-coordinate space
from the action, and $\Sigma$ is redefined as the space spanned by
the $x$-coordinate. Equations of motion from
(\ref{action.reduced}) is consistent with
$\wtd{R}_{\mu\nu}=\frac{2\Lambda}{D-2}\wtd{G}_{\mu\nu}$. When
varying $H_{IJ}$, it is important to note that
%%%
\bea\sqrt{|G|}\; \delta R&=&\sqrt{|G|}\;\Big(R_{IJ}
-\nabla_I\nabla_J +H_{IJ}\nabla^2\Big)\delta H^{IJ}\nn\\
&=&\sqrt{|G|}\;\Big\{R_{IJ}-\frac{\nabla_I\nabla_J
\sqrt{|G|}}{\sqrt{|G|}}+H_{IJ}\frac{\nabla^2
\sqrt{|G|}}{\sqrt{|G|}}\Big\}\delta H^{IJ}\,,\\
&&+{\rm boundary~terms~(to~be~cancelled
~by~the~boundary~action)}\,.\nn\eea
%%%
By tracing over $\wtd{R}_{AB}=\frac{2\Lambda}{D-2}\wtd{G}_{AB}$,
we also find
%%%
\be(\pd\ln \sqrt{|G|})^2+\nabla^2\ln\sqrt{|G|}=\frac{\nabla^2
\sqrt{|G|}}{\sqrt{|G|}} =-\frac{2m\Lambda}{D-2}\,.
\label{nice.identity}\ee
%%%

It is obvious that (\ref{action.reduced}) has a rigid $SL(m,R)$
symmetry: the action is invariant under the transformation,
%%%
\be G_{AB}\quad\longrightarrow\quad(\cV\cdot G\cdot\cV^T)_{AB}
\,,\quad|\cV|=1\,. \label{sl2r}\ee
%%%
This symmetry is due to the freedom in redefining the
$y$-coordinates,
%%%
\be dy^A\quad\longrightarrow\quad (dy\cdot\cV^{-1})^A\,.\ee
%%%
As such, the same symmetry should continue to exist even when
there are additional matter fields. Of course, the matter fields
should transform appropriately to keep the physical objects
invariant. For example, a vector field should transform as
%%%
\be\cA_I\quad\longrightarrow\quad\cA_I\,,\quad
\cA_A=\quad\longrightarrow\quad(\cV\cA)_A\,,\ee
%%%
which leaves $\cA=dx^I\cA_I+dy^A\cA_A$ invariant.

\section{First Law for Stationary and Axisymmetric Black Holes}

It is well known that any metric can be cast into the ADM form,
%%%
\be ds^2=-N^2dt^2+g_{ij}(dx^i-N^idt)(dx^j-N^jdt)\,.\ee
%%%
For a stationary and axisymmetric black hole, the metric elements
are further constrained, and the metrics can
always be put into the following form \cite{mei10},
%%%
\be ds^2=f\Big[-\frac{\Delta}{v^2}dt^2+\frac{dr^2}\Delta\Big]
+h_{ij}d\theta^i d\theta^j+g_{ab}(d\phi^a-w^adt)(d\phi^b-w^bdt)
\,,\label{metric.general}\ee
%%%
where $\Delta=\Delta(r)$, and the functions $f, v, h_{ij}, g_{ab}$
and $w^a$ depend only on the $r$ and $\theta$-coordinates. In
principle, one can identify the coordinates as the asymptotic time
$t$, the radial coordinate $r$, the latitudinal angles $\theta^i$
($i=1,\cdots, [\frac{D}2]-1$) and the azimuthal angles $\phi^a$
($a=1,\cdots, [\frac{D+1}2]-1$), where $D$ is the total dimension
of the spacetime. The black hole horizon $r_0$ is located at the
(largest) root of $\Delta(r_0)=0$. Near the black hole horizon,
$f, v^2, (h_{ij})$ and $(g_{ab})$ are all positive definite. The
fact that black holes are intrinsically regular on the horizon
puts extra constraints on the functions,
%%%
\bea v(r,\theta^i)&=&v_0(r)+v_1(r,\theta^i)\Delta+\cO(\Delta^2)\,,\nn\\
w^a(r,\theta^i)&=&w_0^a(r)+w^a_1(r,\theta^i)\Delta+\cO(\Delta^2)\,,
\label{wa.expansion}\eea
%%%
which means that any dependence of $v$ and $w^a$ on $\theta^i$ can
only begin at the order $\Delta$. What's more, $v_0(r_0)\neq0$ and
$w_0^a(r_0)=\Omega^a$ is the angular velocity of the black hole in
the $\phi^a$ direction. One can also choose the coordinate system
to be non-rotating at the spatial infinity
($r\rightarrow+\infty$), which means that\footnote{As a side
remark, note if we use (\ref{metric.general}) in the construction
of \cite{mei11}, we will get a vector field that interpolates the
null Killing vector on the horizon and the time Killing vector at
the spatial infinity.}
%%%
\be w^a(r,\theta^i)\quad\longrightarrow\quad 0\quad{\rm as}\quad
r\rightarrow+\infty\,.\ee
%%%
The inverse of (\ref{metric.general}) is
%%%
\be (\pd_S)^2=\frac{\Delta}{f}\pd_r^2 +h^{ij} \pd_{\theta^i}
\pd_{\theta^j} +g^{ab}\pd_{\phi^a}\pd_{\phi^b} -\frac{v^2}{f
\Delta} (\pd_t+w^a\pd_{\phi^a})(\pd_t+w^b \pd_{\phi^b})\,.
\label{metric.inverse}\ee
%%%

It is obvious that (\ref{metric.general}) is a special case of
(\ref{metric}). Comparing (\ref{metric.general}) with
(\ref{metric}), we see that $r$ and $\theta^i$'s belong to the
$x$-coordinates and are labelled by the $I,J,K$ indices, while $t$
and $\phi^a$'s belong to the $y$-coordinates and are labelled by
the $A,B,C$ indices. Also $n=[\frac{D}2]$ and $m=[\frac{D+1}2]$.
The non-vanishing elements of the metric are
%%%
\bea H_{rr}=\frac{f}\Delta\,,\quad H_{ij}=h_{ij}\,,\quad G_{at}
=-w_a\,,\quad G_{ab}=g_{ab}\,,\quad G_{tt}=-\frac1\varrho+w^2\,,\nn\\
H^{rr}=\frac\Delta{f}\,,\quad H^{ij}=h^{ij}\,,\quad
G^{at}=-\varrho w^a\,,\quad G^{ab}=g^{ab}-\varrho w^aw^b\,,\quad
G^{tt}=-\varrho\,, \label{elements.metric}\eea
%%%
where $\varrho=\frac{v^2}{f\Delta}$, $w_a=g_{ab}w^b$ and
$w^2=w_aw^a$. For the determinants, we have
$\sqrt{H}=\sqrt{fh/\Delta}$ and $\sqrt{|G|}=\sqrt{g/\varrho}$ ,
with $h$ being the determinant of $h_{ij}$ and $g$ the determinant
of $g_{ab}$. Note $H>0$ outside the black hole horizon. In the
following, we will still denote $H_{rr}$ and $H_{ij}$ collectively
as $H_{IJ}$, $I,J\in\{r,i\}$. The action (\ref{action.reduced})
can now be written as
%%%
\bea S&=&\int_\Sigma(d^{n-1}\theta)dr\,\cL +\int_{\pd\Sigma}
(d^{n-1}\theta\;dr)_I\;n^I\sqrt{H g/\varrho}\,K\,,\nn\\
\cL&=&\sqrt{H g/\varrho}\,\Big\{R-2\Lambda+(\pd\ln
\sqrt{g/\varrho}\,)^2 -(\pd\ln\sqrt\varrho\,)^2\nn\\
&&\qquad\qquad\quad+\frac14\pd g_{ab}\pd g^{ab}
+\frac\varrho2g_{ab}\pd w^a\pd w^b \Big\}\,.\label{action.BH}\eea
%%%
Note this action is completely regular on the black hole horizons
($\Delta\rightarrow0$). This is reasonable because black holes are
intrinsically regular on the horizons. As mentioned before, one
can formally treat (\ref{action.BH}) as a field theory of
$g_{ab}$, $\varrho$ and $w^a$, defined in the curved background
$H_{IJ}$. So correspondingly, one can derive a new set of
equations of motion,
%%%
\bea-\frac{\nabla(\sqrt{g/\varrho}\;\pd g_{ab})}{2\sqrt{
g/\varrho}} +\frac12g^{cd}\pd g_{ac}\pd g_{bd} -\frac\varrho2
g_{ac}g_{bd}\pd w^c\pd w^d&=&\frac{2\Lambda}{D-2}g_{ab}\,,
\label{eom.gab}\\
\frac{\nabla(\sqrt{g/\varrho}\;\pd\ln\sqrt\varrho)}{\sqrt{g/\varrho}}
+\frac\varrho2 g_{ab}\pd w^a\pd w^b&=&\frac{2\Lambda}{D-2}\,,
\label{eom.varrho}\\
\nabla\Big(\sqrt{g/\varrho}\;\varrho g_{ab}\pd w^b\Big)&=&0\,,
\label{eom.wa}\eea
%%%
which are equivalent to $\wtd{R}_{AB}=\frac{2\Lambda}{D-2}
\wtd{G}_{AB}$, $A,B\in\{t,a\}$. By tracing over (\ref{eom.gab})
and then using (\ref{eom.varrho}), we find (note $\delta^a_a=m-1$)
%%%
\be\frac{\nabla^2\sqrt{g/\varrho}}{\sqrt{g/\varrho}} =-\frac{2m
\Lambda}{D-2}\,,\label{nice.identity2}\ee
%%%
thus recovering (\ref{nice.identity}). Also, we can vary $H_{IJ}$
to obtain
%%%
\bea\frac{2\Lambda}{D-2}H_{IJ}&=& R_{IJ}-\nabla_I\nabla_J\ln
\sqrt{g/\varrho}-\pd_I\ln\sqrt\varrho\;\pd_J\ln\sqrt\varrho\nn\\
&&+\frac14\pd_I g_{ab}\pd_J g^{ab}+\frac\varrho2g_{ab}\pd_Iw^a
\pd_Jw^b\,,\label{eom.HIJ}\eea
%%%
which is equivalent to $\wtd{R}_{IJ}=\frac{2\Lambda}{D-2}
\wtd{G}_{IJ}$, $I,J\in\{r,i\}$.

As an application of the new formalism, let's re-derive the first
law of black hole thermodynamics in terms of the new language. To
facilitate our discussion, we firstly recall some basic formulae
of the covariant phase space method, for which we follow
\cite{wald93,iyer.wald94}.

Consider the general action,
%%%
\be S=\int_\cM\bL\,,\quad \bL=\cL(\Phi^a, \pd_\mu\Phi^a,
\pd_\mu\pd_\nu\Phi^a,\cdots)\ast\one\,,\ee
%%%
where $\Phi$ denotes all possible fields collectively. Through out
the paper, we will use a bold faced letter (e.g. $\bL$) to denote
a differential form.\footnote{We will use the notation
%%%
\be(d^{D-p}x)_{\mu_1\cdots\mu_p}\equiv\frac1{p!(D-p)!}
\varepsilon_{\mu_1\cdots\mu_p\nu_1\cdots\nu_{D-p}^{~}}
dx^{\nu_1}\wedge\cdots\wedge dx^{\nu_{D-p}^{~}}\,,\quad
|\varepsilon_{\cdots}|=1\,,\ee
%%%
with which the Hodge-$\ast$ dual of a $p$-form $\bw_p=\frac1{p!}
w_{\mu_1 \cdots \mu_p}dx^{\mu_1}\wedge\cdots\wedge dx^{\mu_p}$ can
be written as
%%%
\be\ast\bw_p=\sqrt{|g|}\;(d^{D-p}x)_{\mu_1\cdots\mu_p}
w^{\mu_1\cdots\mu_p}\,,\quad\Longrightarrow\quad
\ast\one=\sqrt{|g|}\;d^Dx\,.\ee
%%%
For the exterior and interior products, one has
%%%
\bea d\ast\bw_p=\sqrt{|g|}\;(d^{D-p+1}x)_{\mu_1\cdots
\mu_{p-1}}\nabla_{\mu_p}w^{\mu_1\cdots\mu_p}\,,\nn\\
i_\xi(d^{D-p}x)_{\mu_1\cdots\mu_p}=(d^{D-p-1}x)_{\mu_1
\cdots\mu_p\mu}(p+1)\xi^\mu\,.\eea} For an arbitrary variation of
the fields,
%%%
\be\delta\bL=(\delta\Phi^a)E_a\ast\one +d{\bf\Theta}_\delta\,,\ee
%%%
where all the terms involving a derivative on $\delta\Phi^a$ have
been moved into $d\bf\Theta_\delta$. The Euler-Lagrange equations
are just $E_a=0$. For the special case of a general diffeomorphism
$(\delta=\lie_\xi=d\cdot i_\xi +i_\xi\cdot d)$,
%%%
\bea\lie_\xi\bL=d(i_\xi\bL)=(\lie_\xi\Phi^a)E_a\ast\one +d{\bf
\Theta}_\xi\,,\quad\bJ_\xi ={\bf\Theta}_\xi -i_\xi\bL\,,\nn\\
\Longrightarrow\quad d\bJ_\xi=-(\lie_\xi\Phi^a)E_a\ast\one
\approx0\,,\quad\Longrightarrow\quad \bJ_\xi\approx
d\bQ_\xi\,,\label{current.general}\eea
%%%
where $``\approx"$ means equal after using the equations of motion
$E_a=0$. Now lets evolve a classical solution to a nearby one (We
will focus on the particular operation $\bar\delta$ that only
changes free parameters, such as mass and angular momenta, in the
solution),
%%%
\be\bar\delta\bJ_\xi=\bar\delta{\bf\Theta}_\xi -\bar\delta
(i_\xi\bL) =\bar\delta{\bf\Theta}_\xi-i_\xi\cdot d{\bf
\Theta}_{\bar\delta}=\bw(\bar\delta, \lie_\xi)
+d(i_\xi{\bf\Theta}_{\bar\delta})\,, \quad
\bw(\delta,\lie_\xi)\equiv\delta \bQ_\xi -\lie_\xi
\bQ_\delta\,.\ee
%%%
Since $\bar\delta$ only goes through classical solutions, one has
$\bJ_\xi=d\bQ_\xi$ all the time. Hence
%%%
\be\bar\delta\bJ_\xi=d\bar\delta\bQ_\xi\,,\quad\Longrightarrow
\quad \bw(\bar\delta,\lie_\xi)=d\bk(\bar\delta,\lie_\xi)\,,
\quad\bk(\bar\delta,\lie_\xi)\equiv\bar\delta\bQ_\xi -i_\xi
{\bf\Theta}_{\bar\delta}\,.\ee
%%%
In the case when $\xi$ is a Killing vector of some classical
solution,
%%%
\be\lie_\xi=0\quad \Longrightarrow\quad \bw(\bar\delta,
\lie_\xi)=0\,,\quad \Longrightarrow\quad 0=\int_V\bw(\bar
\delta,\lie_\xi)=\oint_{\pd V} \bk(\bar\delta,\lie_\xi)
\,,\label{indentity1}\ee
%%%
where $V$ is a cauchy surface. Since in this paper we are mainly
interested in stationary and axisymmetric black holes
(\ref{metric.general}), we can take $V$ to be the space outside
the horizon(s). As a result, $\pd V$ has two disconnect pieces:
one at the spatial infinity and one at the (outer) horizon,
%%%
\be\oint_{\pd V}=\int_{+\infty}-\int_{Horizon}\,.
\label{indentity2}\ee
%%%
Usually one defines the charge corresponding to $\lie_\xi$ through
an integral at the spatial infinity,
%%%
\be\bar\delta H_\xi=\int_{+\infty} \bk(\bar\delta, \lie_\xi)
=\int_{+\infty}(\bar\delta\bQ_\xi -i_\xi
{\bf\Theta}_{\bar\delta})\,.\ee
%%%
But because of (\ref{indentity1}) and (\ref{indentity2}), this is
equivalent to defining
%%%
\be\bar\delta H_\xi=\int_{horizon} \bk(\bar\delta, \lie_\xi)
=\int_{horizon}(\bar\delta\bQ_\xi -i_\xi
{\bf\Theta}_{\bar\delta})\,. \label{def.Hxi}\ee
%%%
It is this second definition that we want to use in the following.

Now consider Einstein gravity plus a cosmological constant,
%%%
\be \bL=\Big(\frac{\wtd{R} -2\Lambda}{16\pi}\Big)\ast\one\,,
\label{action.16pi}\ee
%%%
where we use $\wtd{G}_{\mu\nu}$ to denote the full metric
(\ref{metric}), with (\ref{metric.general}) being a special case.
Note we have introduced the factor $\frac1{16\pi}$ into the
Lagrangian density, just to be consistent with the usual
convention of defining charges in general relativity. We will keep
this factor only until the end of this section, and starting from
the next section we will go back and use (\ref{action}) again. For
an arbitrary variation of the fields,
%%%
\bea\delta\bL=\frac1{16\pi}\Big\{\frac{\td{h}}2(\wtd{R}-2 \Lambda)
+(-\wtd{R}^{\mu\nu}+\wtd\nabla^\mu\wtd\nabla^\nu-\wtd \nabla^2
\wtd{G}^{\mu\nu})\td{h}_{\mu\nu}\Big\}\ast\one\,,\nn\\
\Longrightarrow\quad E^{\mu\nu}=\frac1{16\pi}\Big[\frac12\wtd{
G}^{\mu \nu} (\wtd{R}-2\Lambda)-\wtd{R}^{\mu \nu}\Big]\,,\nn\\
{\bf\Theta}_{\delta}=\sqrt{-\wtd{G}}\;(d^{D-1}x)_\mu
\Big(\frac{\wtd\nabla_\nu \td{h}^{\mu\nu} -\wtd\nabla^\mu
\td{h}}{16\pi}\Big)\,,\label{Theta.delta}\eea
%%%
where $\td{h}_{\mu\nu}\equiv\delta \wtd{G}_{\mu \nu}$. (Do not
confuse it with the metric elements $h_{ij}$ in
(\ref{metric.general}).) For a diffeomorphism, one has from
(\ref{current.general})
%%%
\bea\bJ_\xi&=&{\bf\Theta}_\xi-i_\xi\bL=\sqrt{-\wtd{G}}\;(d^{D-1}
x)_\mu \Big\{\frac{-\wtd\nabla_\nu\xi^{\mu\nu} +2\wtd{R}^{\mu\nu}
\xi_\nu}{16\pi} -\Big(\frac{\wtd{R}
-2\Lambda}{16\pi}\Big)\xi^\mu\Big\}\nn\\
&=&\sqrt{-\wtd{G}}\;(d^{D-1}x)_\mu\Big(\frac{-\wtd\nabla_\nu
\xi^{\mu\nu}}{16\pi} \Big)=d\bQ_\xi\,,\nn\\
&\Longrightarrow&\bQ_\xi=\sqrt{-\wtd{G}}\; (d^{D-2}x)_{\mu\nu}
\Big(\frac{-\xi^{\mu\nu}}{16\pi}\Big)\,,\quad \xi^{\mu\nu}
=\wtd\nabla^\mu\xi^\nu -\wtd\nabla^\nu \xi^\mu\,.\eea
%%%
The metric (\ref{metric.general}) has the Killing vectors
$\hat{k}=\pd_t$ and $\hat{k}_a =\pd_{\phi^a}$. The elements
relevant for the integral (\ref{def.Hxi}) are
%%%
\bea \hat{k}^{tr}&=&\wtd{G}^{t\mu}\wtd{G}^{rr}(\pd_\mu
\hat{k}_r-\pd_r\hat{k}_\mu) =-\wtd{G}^{t\mu}
\wtd{G}^{rr}\pd_r\wtd{G}_{t\mu}=-\frac\Delta{f}
\varrho\Big[\pd_r(\frac1\varrho-w^2) +w^a\pd_rw_a\Big]\nn\\
&=&\frac{v^2}{f^2}\Big(\frac1\varrho\pd_r\ln\varrho +w_a\pd_r
w^a\Big)=\frac{v^2}{f^2}\Big(w_a\pd_rw^a-\frac{f\Delta'}{v^2}
+\frac{f\Delta}{v^2}\pd_r\ln\frac{v^2}f\Big)\,,\nn\\
&\longrightarrow&\frac{v^2}{f^2}\Big(w_a\pd_rw^a
-\frac{f\Delta'}{v^2}\Big)\,,\label{def.ktr}\\
\hat{k}_a^{tr}&=&\wtd{G}^{t\mu}\wtd{G}^{rr}\Big[\pd_\mu
(\hat{k}_a)_r -\pd_r (\hat{k}_a)_\mu\Big]=-\wtd{G}^{t\mu}
\wtd{G}^{rr}\pd_r\wtd{G}_{a\mu}\nn\\
&=&-\frac\Delta{f}\varrho\Big[\pd_rw_a-w^b\pd_rg_{ab}\Big]
=-\frac{v^2}{f^2}g_{ab}\pd_rw^b\,,\eea
%%%
where $``\longrightarrow"$ means equal in the limit $\Delta
\rightarrow0$. Similarly using (\ref{Theta.delta}), one has for
$i_\xi{\bf\Theta}_{\bar\delta} =\sqrt{-\wtd{G}} (d^{D-2}x)_{\mu
\nu}(i_\xi{\bf\Theta}_{\bar \delta})^{\mu\nu}$,
%%%
\bea(i_\xi{\bf\Theta}_{\bar\delta})^{\mu\nu}&=&\xi^\nu\Big(
\frac{\wtd\nabla_\rho\bar{h}^{\mu\rho}-\wtd\nabla^\mu \bar{h}}{16
\pi}\Big) -\xi^\mu\Big(\frac{ \wtd\nabla_\rho\bar{h}^{\nu\rho}
-\wtd\nabla^\nu \bar{h}}{16\pi}\Big)\,,\nn\\
(i_{\hat{k}}{\bf\Theta}_{\bar\delta})^{tr}&=&-\frac1{16\pi}
\Big(\wtd\nabla_\mu\bar{h}^{r\mu}-\wtd\nabla^r\bar{h}\Big)\nn\\
&=&-\frac1{16\pi}\Big(\pd_r\bar{h }^{rr} +\td\Gamma^r_{\mu\nu}
\bar{h}^{\mu\nu}+\td\Gamma^\mu_{\mu r} \bar{h}^{rr} -\wtd{G}^{rr}
\pd_r\bar{h}\Big)\nn\\
&=&-\frac1{16\pi}\Big(\pd_r \bar{h}^{rr}
+\wtd{G}^{rr}\pd_r\wtd{G}_{rr} \bar{h}^{rr}
-\frac12\wtd{G}^{rr}\pd_r \wtd{G}_{\mu\nu}
\bar{h}^{\mu\nu} +\bar{h}^{rr}\pd_r\ln\sqrt{-\wtd{G}}\nn\\
&&\qquad\quad-2\wtd{G}^{rr}\pd_r\bar\delta\ln\sqrt{-\wtd{G}}\;\Big)\nn\\
&\longrightarrow&-\frac1{16\pi}\Big(\pd_r \bar{h}^{rr}
+\wtd{G}^{rr} \pd_r\wtd{G}_{rr} \bar{h}^{rr}
+\frac12\wtd{G}^{rr}\pd_r\wtd{G}^{\mu\nu}
\bar{h}_{\mu\nu}+\bar{h}^{rr}\pd_r\ln\sqrt{-\wtd{G}}\;\Big)\nn\\
&\longrightarrow&-\frac1{16\pi}\Big\{\pd_r\Big(\frac{\Delta^2}{
f^2} \bar\delta\frac{f}\Delta\Big) +\frac{\Delta^2}{f^2} \bar
\delta \frac{f}\Delta\pd_r\ln\frac{f}\Delta +\frac\Delta{2f}
\Big[\pd_r\frac\Delta f\bar\delta \frac{f}\Delta +\pd_r\varrho
\bar\delta(\frac1\varrho -w^2)\nn\\
&&\qquad\quad +2\pd_r(\varrho w^a)\bar\delta w_a +\pd_r(g^{ab}
-\varrho w^aw^b) \bar\delta g_{ab}\Big]\Big\}\nn\\
&\longrightarrow&-\frac1{16\pi}\Big[\frac{v^2}{f^2} g_{ab}\pd_r
w^a \bar\delta w^b-\frac{v}f\bar\delta\Big(\frac{\Delta'}v\Big)
\Big]\,,\eea
%%%
where $\bar{h}_{\mu\nu}\equiv\bar\delta\wtd{G}_{\mu \nu}$. (Do not
confuse it with the metric elements $h_{ij}$ in
(\ref{metric.general}).) Note although we have kept $\Delta$
explicit (at where it is necessary) to show that none of the
expressions diverge in the limit $\Delta\rightarrow0$, it should
be understood that the operation $\bar\delta$ always comes {\it
after} taking the limit $r\rightarrow r_0$. For this reason,
$\bar\delta\Delta=0$ holds all the time. Plugging the results back
into (\ref{def.Hxi}), we find
%%%
\bea\bar\delta E&=&\bar\delta H_{\hat{k}}=\int_{r=r_0} (d^{D-2}
x)_{\mu\nu}\Big\{\bar\delta\Big(\sqrt{-\wtd{G}}\; \frac{-\hat{
k}^{\mu \nu}}{16\pi}\Big) -\sqrt{-\wtd{G}}\; (i_{\hat{k}}
{\bf\Theta}_{\bar\delta})^{\mu\nu}\Big\}\nn\\
&=&\int_{r=r_0} (d^{D-2} x)_{tr}2\Big\{\bar\delta \Big(-\frac{
\sqrt{h g}}{16\pi}\frac{v}f w_a\pd_r w^a +\frac{\sqrt{h
g}}{16\pi}\frac{\Delta'}v\Big) \nn\\
&&\qquad\qquad\qquad\qquad+\frac{\sqrt{h g}}{16\pi}\frac{v}f
g_{ab}\pd_r w^a\bar\delta w^b-\frac{\sqrt{h g}}{16\pi}\bar\delta
\Big(\frac{\Delta'}v\Big)\Big\}\nn\\
&=&\int_{r=r_0}(d^{D-2}x)_{tr}2\Big\{w^a\bar\delta \Big(-\frac{
\sqrt{h g}}{16\pi}\frac{v}f g_{ab}\pd_rw^b\Big)+\frac{\Delta'}{16
\pi v}\bar\delta\sqrt{h g}\;\Big\}\nn\\
&=&T\bar\delta S +\Omega^a\bar\delta J_a\,, \label{first.law}\eea
%%%
where we have used $\sqrt{-\wtd{G}}=\sqrt{h g}\ds\frac{f}v$ and in
the last step the definitions
%%%
\bea T&=&\frac\kappa{2\pi}=\frac{\Delta'}{4\pi v}
\Big|_{r=r_0}\,,\quad \Omega^a=w^a(r_0)\,,\nn\\
J_a&=&-H_{\hat{k}_a} =\int_{r=r_0}\sqrt{-\wtd{G}}\;
(d^{D-2}x)_{\mu\nu}\Big(\frac{\hat{k}_a^{\mu\nu}}{16\pi}\Big)\nn\\
&=&\int_{r=r_0}(d^{D-2}x)_{tr}2\Big(-\frac{ \sqrt{h g}}{16\pi}
\frac{v}f g_{ab}\pd_rw^b\Big)\,,\nn\\
S&=&\frac14\int_{r=r_0}(d^{D-2}x)_{tr}2\sqrt{h
g}=\frac{\cA_{rea}}4\;,\label{charges}\eea
%%%
where $\kappa$ is the surface gravity on the horizon.

Note the above calculation is not a true ``derivation" of the
first law because the $\bar\delta$-integrability of
(\ref{def.Hxi}) is not {\it a priori} obvious. As such, the above
calculation, together with the observation that one can integrate
the first law to recover the black hole masses
\cite{gibbons.perry.pope04}, can be better interpreted as showing
that (\ref{def.Hxi}) is $\bar\delta$-integrable for stationary and
axisymmetric black holes, in the context of Einstein gravity plus
a cosmological constant.

As a side remark, note \cite{wald93,iyer.wald94} already involved
deriving the first law of thermodynamics from the general calculus
of the covariant phase space method. What's new here is that (i)
we are using an operation $\bar\delta$ that is directly related to
the usual test of the first law of black hole thermodynamics, and
(ii) all the quantities are now defined at the black hole horizon,
without any reference to the spatial infinity (But because of
(\ref{indentity1}) and (\ref{indentity2}), the results must be the
same).

We want to emphasize that the above calculation becomes possible
only because our formalism has made the dependence on the function
$\Delta(r)$ explicit, which holds key informations of the metric
(\ref{metric.general}) as it approaches the black hole horizon.

\section{The Conformal Symmetries on the Horizon}

As was mentioned before, the action (\ref{action.reduced}) has a
rigid $SL(m,R)$ symmetry, which should be inherited by the
particular case (\ref{action.BH}). In this section, we want to
focus on the particular $SL(2,R)$ generators like the following,
%%%
\be L_0=\frac12\left(\begin{matrix}-1&\cdots&0\cr\vdots&
\ddots&\vdots&\cr0&\cdots&1\end{matrix}\right)\,,\quad
L_+=\left(\begin{matrix}0&\cdots&1\cr \vdots&\ddots&
\vdots&\cr0&\cdots&0\end{matrix}\right)\,,\quad
L_-=\left(\begin{matrix}0&\cdots&0\cr \vdots&\ddots&
\vdots&\cr-1&\cdots&0\end{matrix}\right)\,,\label{sl2r.g2}\ee
%%%
where all the matrices are $m$-dimensional, and all the implicit
elements are zero. The transformation of the metric elements
$G_{AB}$ will be given by
%%%
\be\hat\delta G\equiv-(L\cdot G+G\cdot L^T)\,.
\label{transform.sl2r}\ee
%%%
In order to see the results explicitly, let's distinguish the
coordinate $\phi^1$ from the rest of the azimuthal angles. We will
simply denote $\phi^1$ as $\phi$, and will also use $\phi$ as the
corresponding super/sub-script, e.g. $w^1=w^\phi$ and
$g_{11}=g_{\phi\phi}$. We will label all other azimuthal angles
using indices with a tilde, $\phi^\tda$ ($\tda=2,\cdots,m-1$).
Accordingly,
%%%
\bea (G_{AB})&=&\left(\begin{matrix}g_{\phi\phi}&g_{\tda\phi}
&-w_\phi\cr g_{\tdb\phi}&g_{\tda\tdb}&-w_\tdb \cr-w_\phi&-w_\tda
&-\frac1 \varrho+w^2\end{matrix}\right)\,,\nn\\
(G^{AB})&=&\left(\begin{matrix}g^{\phi\phi}-\varrho w^\phi w^\phi
&g^{\tda\phi}-\varrho w^\tda w^\phi&-\varrho w^\phi\cr
g^{\tdb\phi}-\varrho w^\tdb w^\phi&g^{\tda\tdb}-\varrho w^\tda
w^\tdb &-\varrho w^\tdb \cr-\varrho w^\phi&-\varrho w^\tda
&-\varrho\end{matrix}\right)\,.\eea
%%%
Note both the indices $\{\phi,\td{a}\}$ are still raised and
lowered using the matrix
%%%
\be (g_{ab})=\left(\begin{matrix}g_{\phi\phi}&g_{\tda\phi}\cr
g_{\tdb\phi}&g_{\tda\tdb}\end{matrix}\right)\,,\quad
(g^{ab})=\left(\begin{matrix}g^{\phi\phi}&g^{\tda\phi}\cr
g^{\tdb\phi}&g^{\tda\tdb}\end{matrix}\right)\,.\ee
%%%
As such, we will try to convert our results back to using the
untilded indices (which take values form $\{\phi,2,\cdots,m-1\}$)
whenever it is possible.

Our following construction will also rely on the assumption that
$w^\phi=w^1\neq0$. But the choice on $\phi^1$ is only a matter of
convenience. One can do the same for any other azimuthal angles,
as long as the corresponding angular velocity is non-zero. Of
course, one should accordingly relocate the the non-vanishing
matrix elements in (\ref{sl2r.g2}).

Now using (\ref{transform.sl2r}), we find for the symmetric
transformations,
%%%
\bea\hat\delta_0g_{\phi\phi}&=&g_{\phi\phi}\,,\qquad\quad
\hat\delta_0g^{\phi\phi}=-g^{\phi\phi}\,,\nn\\
\hat\delta_0g_{\tda\phi}&=&\frac12g_{\tda\phi}\,,\qquad~
\hat\delta_0g^{\tda\phi}=-\frac12g^{\tda\phi}\,,\nn\\
\hat\delta_0g_{\tda\tdb}&=&0\,,\qquad\qquad
\hat\delta_0g^{\tda\tdb}=0\,,\nn\\
\hat\delta_0w^\phi&=&-w^\phi\,,\qquad~ \hat\delta_0w_\phi=0\,,\nn\\
\hat\delta_0w^\tda &=&-\frac12w^\tda \,,\quad~~ \hat\delta_0w_\tda
=-\frac12w_\tda  \,,\quad
\hat\delta_0\varrho=\varrho\,,\label{delta0.g2}\\
&&----------------\nn\\
\hat\delta_+g_{\phi\phi}&=&2w_\phi\,,\qquad\qquad\qquad\quad~
\hat\delta_+g^{\phi\phi}=-2g^{\phi\phi}w^\phi\,,\nn\\
\hat\delta_+g_{\tda\phi}&=&w_\tda \,,\qquad\qquad\qquad\qquad
\hat\delta_+g^{\tda\phi}=-(g^{\tda\phi}w^\phi+g^{\phi\phi}w^\tda )\,,\nn\\
\hat\delta_+g_{\tda\tdb}&=&0\,,\qquad\qquad\qquad\qquad~~
\hat\delta_+g^{\tda\tdb}=-(g^{\tda\phi}w^\tdb +g^{\tdb\phi}w^\tda )\,,\nn\\
\hat\delta_+w^\phi&=&-(w^\phi w^\phi+g^{\phi\phi}/\varrho)
\,,\quad~\hat\delta_+w_\phi=-\frac1\varrho +w^2\,,\nn\\
\hat\delta_+w^\tda &=&-(w^\tda
w^\phi+g^{\tda\phi}/\varrho)\,,\quad~ \hat\delta_+w_\tda
=0\,,\qquad\hat\delta_+\varrho =2\varrho
w^\phi\,,\label{delta1p.g2}\\
&&----------------\nn\\
\hat\delta_-g_{\phi\phi}&=&0\,,\qquad
\hat\delta_-g^{\phi\phi}=0\,,\nn\\
\hat\delta_-g_{\tda\phi}&=&0\,,\qquad
\hat\delta_-g^{\tda\phi}=0\,,\nn\\
\hat\delta_-g_{\tda\tdb}&=&0\,,\qquad
\hat\delta_-g^{\tda\tdb}=0\,,\nn\\
\hat\delta_-w^\phi&=&-1\,,\quad~
\hat\delta_-w_\phi=-g_{\phi\phi}\,,\nn\\
\hat\delta_-w^\tda &=&0\,,\qquad \hat\delta_-w_\tda =-g_{\tda\phi}
\,,\quad\hat\delta_-\varrho=0\,,\label{delta1n.g2}\eea
%%%
It is easy to check that
%%%
\be[\hat\delta_\pm\,,\,\hat\delta_0]=\pm\hat\delta_\pm\,,\quad
~[\hat\delta_+\,,\,\hat\delta_-]=2\hat\delta_0\,.
\label{algebra.sl2r.g2}\ee
%%%
For later convenience, lets define
%%%
\bea\pi^{Iab}&=&\frac{\delta S}{\delta(\pd_Ig_{ab})} =\sqrt{H
g/\varrho}\; \Big(g^{ab}\pd^I\ln\sqrt{g/\varrho}\;+\frac12
\pd^Ig^{ab}\Big)\,,\nn\\
\pi^I_a&=&\frac{\delta S}{\delta(\pd_Iw^a)} =\sqrt{H
g/\varrho}\; \Big(\varrho g_{ab}\pd^Iw^b\Big)\,,\nn\\
\pi^I_\varrho&=&\frac{\delta S}{\delta(\pd_I\varrho)} =\sqrt{H
g/\varrho}\; \Big(-\frac1\varrho\pd^I\ln\sqrt{g}\Big)\,.
\label{def.pi}\eea
%%%
The Noether currents corresponding to (\ref{delta0.g2}),
(\ref{delta1p.g2}) and (\ref{delta1n.g2}) are
%%%
\bea J_0^I&=&\pi^{Iab} \hat\delta_0 g_{ab} +\pi^I_a\hat
\delta_0w^a+\pi^I_\varrho\hat\delta_0\varrho\,,\nn\\
&=&\sqrt{H g/\varrho}\;\Big(\frac12g_{\phi a}\pd^Ig^{a
\phi}-\pd^I\ln\sqrt\varrho -\frac\varrho2w^ag_{ab}\pd^I
w^b-\frac\varrho2w^\phi g_{\phi a}\pd^Iw^a\Big)\,,\nn\\
J_+^I&=&\pi^{Iab} \hat\delta_+ g_{ab} +\pi^I_a\hat\delta_+
w^a+\pi^I_\varrho\hat\delta_+\varrho\,,\nn\\
&=&\sqrt{H g/\varrho}\;\Big(-2w^\phi\pd^I\ln\sqrt\varrho
+w_a\pd^Ig^{a\phi}-\pd^Iw^\phi-\varrho w^\phi w_a\pd^I
w^a\Big)\,,\nn\\
J_-^I&=&\pi^{Iab} \hat\delta_- g_{ab} +\pi^I_a\hat\delta_-
w^a+\pi^I_\varrho\hat\delta_-\varrho=\sqrt{H g/\varrho}\;
\Big(-\varrho g_{\phi a}\pd^Iw^a\Big)\,.\label{currents.g2}\eea
%%%
By using the equations of motion (\ref{eom.gab}),
(\ref{eom.varrho}) and (\ref{eom.wa}), one can check that all
these currents are exactly conserved.

There is an interesting connection between these currents and the
charges defined in (\ref{first.law}) and (\ref{charges}). Using
the detail of the metric elements (\ref{elements.metric}) and the
relations $\varrho=\frac{v^2}{f\Delta}$ and $\sqrt{H
g/\varrho}\;=\sqrt{h g}\frac{f}v$, one can find that
%%%
\be J_-^r=\sqrt{h g}\frac{v}f\Big(-g_{\phi a}\pd_rw^a\Big)\,.\ee
%%%
It is obvious that $J_-^r$ is just the integrand of the angular
momentum $J_\phi$ in (\ref{charges}).\footnote{The extra factor
$\frac1{16\pi}$ comes from the difference between (\ref{action})
and (\ref{action.16pi}).} For the energy $E$, it is easier to look
at the asymptotically flat case ($\Lambda=0$). In this case, it is
possible to define the energy as a Komar integral,
%%%
\be E\sim-\int_{+\infty}\ast d\hat{k} =-\int_{Horizon}\ast
d\hat{k}=\int_{Horizon}(d^{D-2}x)_{tr}\sqrt{h g}\frac{f}v
2(-\hat{k}^{tr})\,,\label{energy.komar}\ee
%%%
where $\hat{k}^{tr}$ has been given in (\ref{def.ktr}), and in the
second step we have used $\wtd{R}_{\mu\nu}\sim\Lambda \wtd{G}_{\mu
\nu}=0$ and the relation $\wtd\nabla_\nu \wtd\nabla^\mu \xi^\nu
=\wtd{R}^\mu_{~\nu}\xi^\nu$ which is valid for any Killing vector
$\xi$. Now notice that for each azimuthal angle $\phi^a$, it is
possible to construct a copy of the currents (\ref{currents.g2}).
Using (\ref{nice.identity2}), we see that the following current
(from summing over the $J_0^I$ corresponding to each azimuthal
angles and then subtract out a trivial piece) is also conserved
when $\Lambda=0$,
%%%
\bea J^I&=&\frac2m \sum_{\phi=1}^{m-1} J_0^I +\frac2m\sqrt{H
g/\varrho}\pd^I\ln\sqrt{g/\varrho}\nn\\
&=&-\sqrt{h g}\frac{f}v\Big(\varrho w_a\pd^Iw^a
+2\pd^I\ln\sqrt\varrho\Big)\,,\nn\\
\Longrightarrow\quad J^r&\longrightarrow&-\sqrt{h g}\frac{f}v
\Big(\frac{v^2}{f^2} w_a\pd_rw^a -\frac{2\Delta'}f\Big)\,,\eea
%%%
where $``\longrightarrow"$ means equal in the limit
$\Delta\rightarrow0$. By comparing with (\ref{def.ktr}), we see
that $J^r$ is just the integrand of (\ref{energy.komar}), up to a
normalization constant. Despite the fact that the connections
found in this paragraph is very interesting, they will have
nothing to do with our following discussions.

Given the above $SL(2,R)$ symmetry (\ref{algebra.sl2r.g2}), it is
natural to ask if one can extend it to the infinite dimensional
Witt algebra,
%%%
\be\Big[\hat\delta_\bm\,,\,\hat\delta_\bn\Big]=(\bm -\bn)
\hat\delta_{\bm+\bn}\,,\quad \bm,\bn=0,\pm1,\pm2,\cdots\,.
\label{algebra.witt}\ee
%%%
In particular, we want to see if we can construct operators that
satisfy (\ref{algebra.witt}) approximately near the black hole
horizons, where $\Delta\rightarrow0$ (i.e. $\rho\rightarrow
+\infty$). Technically, given $\hat\delta_{0,\pm}$, one only needs
to figure out $\hat\delta_2$ and $\hat\delta_{-2}$ to obtain the
full algebra: all other operators can then be constructed by
iterating the following relations,
%%%
\be\hat\delta_{\bm+1}=\frac1{\bm-1}\Big[\hat\delta_\bm\,,
\,\hat\delta_+\Big]\,,\quad \hat\delta_{-\bm-1}=\frac1{-\bm+1}
\Big[\hat\delta_{-\bm}\,,\,\hat\delta_-\Big]\,,\quad\bm\geq2\,.
\label{witt.generating}\ee
%%%
We will want all the new transformations $\hat\delta_\bm$
($\bm=\pm2,\pm3,\cdots$) to be regular and non-trivial on the
horizon, just as $\hat\delta_0$ and $\hat\delta_\pm$ in
(\ref{delta0.g2}), (\ref{delta1p.g2}) and (\ref{delta1n.g2}).

To generalize (\ref{delta0.g2}), (\ref{delta1p.g2}) and
(\ref{delta1n.g2}) to infinite dimensions, let's start with
%%%
\bea\Big[\hat\delta_2\,,\,\hat\delta_0\Big]=2\hat\delta_2\,,\quad
\Big[\hat\delta_{-2}\,,\,\hat\delta_0\Big]=-2\hat\delta_{-2}\,.\eea
%%%
In combination with (\ref{delta0.g2}), we find
%%%
\bea\hat\delta_0\hat\delta_2g_{\phi\phi}&=&-\hat\delta_2g_{\phi\phi}\,,\qquad\quad
\hat\delta_0\hat\delta_{-2}g_{\phi\phi}=3\hat\delta_{-2}g_{\phi\phi}\,,\nn\\
\hat\delta_0\hat\delta_2g_{\tda\phi}&=&-\frac32\hat\delta_2g_{\tda\phi}\,,\qquad~
\hat\delta_0\hat\delta_{-2}g_{\tda\phi}=\frac52\hat\delta_{-2}g_{\tda\phi}\,,\nn\\
\hat\delta_0\hat\delta_2g_{\tda\tdb}&=&-2\hat\delta_2g_{\tda\tdb}\,,\qquad~~
\hat\delta_0\hat\delta_{-2}g_{\tda\tdb}=2\hat\delta_{-2}g_{\tda\tdb}\,,\nn\\
\hat\delta_0\hat\delta_2w^\phi&=&-3\hat\delta_2w^\phi\,,\qquad~~
\hat\delta_0\hat\delta_{-2}w^\phi=\hat\delta_{-2}w^\phi\,,\nn\\
\hat\delta_0\hat\delta_2w^\tda&=&-\frac52\hat\delta_2w^\tda\,,\qquad~~
\hat\delta_0\hat\delta_{-2}w^\tda =\frac32\hat\delta_{-2}w^\tda \,,\nn\\
\hat\delta_0\hat\delta_2\varrho&=&-\hat\delta_2\varrho\,,\qquad\qquad
\hat\delta_0\hat\delta_{-2}\varrho=3\hat\delta_{-2}\varrho\,.
\label{dim.sl2r.g2}\eea
%%%
Keeping in mind that $\hat\delta_{\pm2}$ should be regular and
non-trivial on the horizon, and also guided by
(\ref{dim.sl2r.g2}), we try the following ansatz
%%%
\bea\hat\delta_2g_{\phi\phi}&=&u_1g_{\tda\tdb}w^\tda w^\tdb +u_2
g_{\tda\phi}w^\tda w^\phi+u_3g_{\phi\phi}w^\phi w^\phi\,,\nn\\
\hat\delta_2g_{\tda\phi}&=&u_4g_{\tda\tdb}w^\tdb w^\phi +u_5
g_{\tda\phi}w^\phi w^\phi\,,\quad \hat\delta_2g_{\tda\tdb}=0\,,\nn\\
\hat\delta_2w^\phi&=&u_6w^\phi w^\phi w^\phi\,,\quad
\hat\delta_2w^\tda =u_7w^\tda  w^\phi w^\phi\,,\label{delta2p.g2}\\
\hat\delta_{-2}g_{\phi\phi}&=&\frac{v_1g_{\tda\tdb}w^\tda w^\tdb
+v_2 g_{\tda\phi}w^\tda w^\phi+v_3g_{\phi\phi}w^\phi
w^\phi}{w^\phi w^\phi w^\phi w^\phi}\,,\nn\\
\hat\delta_{-2}g_{\tda\phi}&=&\frac{v_4g_{\tda\tdb}w^\tdb  +v_5
g_{\tda\phi} w^\phi}{w^\phi w^\phi w^\phi}\,,\quad\hat\delta_{-2}
g_{\tda\tdb}=0\,,\nn\\
\hat\delta_{-2}w^\phi&=&v_6/w^\phi\,,\quad \hat\delta_{-2}w^\tda
=v_7w^\tda /(w^\phi w^\phi) \,,\label{delta2n.g2}\eea
%%%
where $u_1\,,\cdots\,,u_7$ and $v_1\,,\cdots\,,v_7$ are constants.
Note $g/\varrho$ is invariant under (\ref{delta0.g2}),
(\ref{delta1p.g2}) and (\ref{delta1n.g2}). Here we further assume
that $g/\varrho$ is neutral under all the transformations. This
requirement fully determines the structure of
$\hat\delta_\bm\varrho$ :
%%%
\be\delta_\bm\varrho=\varrho g^{ab}\delta_\bm g_{ab}\,,\quad
\forall \;\bm=0,\pm1,\pm2,\cdots\,.\ee
%%%
Now since (Here ``$\approx$" means equal at the leading order in
$\varrho\rightarrow+\infty$)
%%%
\be\Big[\hat\delta_2\,,\,\hat\delta_-\Big]\approx3\hat\delta_+\,,\quad
\Big[\hat\delta_{-2}\,,\,\hat\delta_+\Big]\approx-3\hat\delta_-\,,\quad
\Big[\hat\delta_2\,,\,\hat\delta_{-2}\Big]\approx4\hat\delta_0\,,
\label{algebra.delta2}\ee
%%%
we find $v_1=-u_1$, and
%%%
\bea u_2=6\,,\quad u_3=3\,,\quad u_4=3\,,\quad u_5=\frac32\,,\quad
u_6=-1\,,\quad u_7=-\frac32\,,\nn\\
v_2=2\,,\quad v_3=-1\,,\quad v_4=1\,,\quad v_5=-\frac12\,,\quad
v_6=-1\,,\quad v_7=\frac12\,.\eea
%%%
The currents corresponding to (\ref{delta2p.g2}) and
(\ref{delta2n.g2}) are
%%%
\bea J_2^I&=&\pi^{Iab}\hat\delta_2 g_{ab}+\pi^I_a\hat
\delta_2w^a+\pi^I_\varrho\hat\delta_2\varrho\,,\nn\\
&=&\sqrt{H g/\varrho}\;\Big\{u_1g_{\tda\tdb}w^\tda w^\tdb
\Big(\frac12\pd^I g^{\phi\phi}-g^{\phi\phi}\pd^I\ln\sqrt
\varrho\Big)-3\pd^I\ln\sqrt\varrho\;w^\phi w^\phi\nn\\
&&\qquad\qquad-3\pd^Ig_{ab}g^{a\phi}w^bw^\phi+\frac32
\pd^Ig_{a\phi} g^{a\phi}w^\phi w^\phi\nn\\
&&\qquad\qquad-\frac32\varrho g_{ab}w^aw^\phi w^\phi
\pd^Iw^b+\frac12\varrho g_{a\phi}w^\phi w^\phi w^\phi
\pd^Iw^a\Big\}\,, \label{current.j2p}\\
J_{-2}^I&=&\pi^{Iab}\hat\delta_{-2} g_{ab}+\pi^I_a\hat
\delta_{-2}w^a+\pi^I_\varrho\hat\delta_{-2}\varrho\,,\nn\\
&=&\frac{\sqrt{H g/\varrho}}{w^\phi w^\phi w^\phi
w^\phi}\Big\{-u_1g_{\tda\tdb}w^\tda w^\tdb \Big(\frac12
\pd^Ig^{\phi\phi}-g^{\phi\phi}\pd^I\ln\sqrt \varrho\Big)
+\pd^I\ln\sqrt\varrho\;w^\phi w^\phi\nn\\
&&\qquad\qquad\quad-\pd^Ig_{ab}g^{a\phi}w^bw^\phi+\frac32
\pd^Ig_{a\phi} g^{a\phi}w^\phi w^\phi\nn\\
&&\qquad\qquad\quad+\frac12\varrho g_{ab}w^aw^\phi w^\phi
\pd^Iw^b-\frac32\varrho g_{a\phi}w^\phi w^\phi w^\phi
\pd^Iw^a\Big\}\,.\label{current.j2n}\eea
%%%
With the help of the equations of motion (\ref{eom.gab}),
(\ref{eom.varrho}) and (\ref{eom.wa}), and also the properties
(\ref{wa.expansion}) and (\ref{elements.metric}), the total
divergence of the currents can be found as
%%%
\bea\pd_IJ_2^I&=&\sqrt{H g/\varrho}\;\Big(-6\pd\ln \sqrt\varrho\;
w^\phi\pd w^\phi\Big)+\cO(\frac1\varrho)+u_1{\rm~term}\,,\nn\\
&=&\sqrt{h g}\;\Big(3\frac{\Delta'}vw^\phi\pd_rw^\phi\Big)
+\cO(\Delta)+u_1{\rm~term}\,,\nn\\
\pd_IJ_{-2}^I&=&\sqrt{H g/\varrho}\;\Big(-2\frac{\pd\ln
\sqrt\varrho\; \pd w^\phi}{w^\phi w^\phi w^\phi}\Big)
+\cO(\frac1\varrho)+u_1{\rm~term}\,,\nn\\
&=&\sqrt{h g}\;\Big(\frac{\Delta'}v\cdot \frac{\pd_r
w^\phi}{w^\phi w^\phi w^\phi}\Big)+\cO(\Delta)
+u_1{\rm~term}\,,\label{divergence.j2}\eea
%%%
where all the $u_1$ terms have components sharing the following
factor,
%%%
\be\pd\ln \sqrt\varrho\;\pd(g_{\tda\tdb}g^{\phi\phi})
=-\frac{\Delta'}{2f}\pd_r(g_{\tda\tdb}g^{\phi\phi})
+h^{ij}\pd_i\ln\sqrt\varrho\;\pd_j(g_{\tda\tdb}
g^{\phi\phi})+\cO(\Delta)\,.\ee
%%%
It is obvious that $\hat\delta_{\pm2}$ are exact symmetries of the
action (\ref{action.BH}) only when both $\Delta'$ and $u_1$ are
zero. We are free to take $u_1=0$ because it is just a
undetermined parameter. On the other hand, $\Delta'$ is related to
the black hole temperature (\ref{charges}), and so it is non-zero
in general. So it appears that the extended symmetries
$\hat\delta_{\pm2}$ are explicitly broken by the finite black hole
temperature.

This problem can be fixed by introducing sub-leading terms into
(\ref{delta2p.g2}) and (\ref{delta2n.g2}),
%%%
\bea&&\hat\delta_2g_{\phi\phi}=6 g_{\tda\phi}w^\tda w^\phi
+3g_{\phi\phi}w^\phi w^\phi+\frac6\varrho\Big(g_{\phi\phi}
g^{\phi\phi}-1\Big)\,,\nn\\
&&\hat\delta_2g_{\tda\phi}=3g_{\tda\tdb}w^\tdb w^\phi+\frac32
g_{\tda\phi}w^\phi w^\phi+3g_{\tda\phi}g^{\phi\phi}/\varrho
\,,\quad\hat\delta_2g_{\tda\tdb}=0\,,\nn\\
&&\hat\delta_2w^\phi=-w^\phi w^\phi w^\phi-3g^{\phi\phi}w^\phi
/\varrho\,,\quad \hat\delta_2w^\tda =-\frac32 w^\tda w^\phi
w^\phi-3 g^{\tda\phi}w^\phi/\varrho\,,\label{delta2p.new}\\
&&\hat\delta_{-2}g_{\phi\phi}=\frac{2g_{\tda\phi}w^\tda w^\phi
-g_{\phi\phi}w^\phi w^\phi - 6(g_{\phi\phi}g^{\phi\phi}
-1)/\varrho}{w^\phi w^\phi w^\phi w^\phi}\,,\nn\\
&&\hat\delta_{-2}g_{\tda\phi}=\frac{g_{\tda\tdb}w^\tdb w^\phi
-\frac12g_{\tda\phi}w^\phi w^\phi-3g_{\tda\phi}g^{\phi
\phi}/\varrho}{w^\phi w^\phi w^\phi w^\phi}\,,
\quad\hat\delta_{-2}g_{\tda\tdb}=0\,,\nn\\
&&\hat\delta_{-2}w^\phi=\frac{-w^\phi w^\phi -g^{\phi\phi}
/\varrho}{w^\phi w^\phi w^\phi}\,,\quad\hat\delta_{-2}w^\tda
=\frac{\frac12w^\tda w^\phi-g^{\tda\phi} /\varrho}{w^\phi w^\phi
w^\phi}\,, \label{delta2n.new}\eea
%%%
where the terms containing $1/\varrho\propto\Delta$ are of the
sub-leading order. The coefficients for each sub-leading terms are
determined by requiring that (\ref{delta2p.new}) and
(\ref{delta2n.new}) satisfy (\ref{algebra.delta2}) up to the
sub-leading order $\cO(\frac1\varrho)$, and also that the currents
are conserved up to $\cO(1)$,
%%%
\bea J_2^I&=&\sqrt{H g/\varrho}\;\Big\{J_\varrho^I-3
w^\phi\pd^Iw^\phi-3\pd^I\ln\sqrt\varrho\;w^\phi w^\phi\nn\\
&&\qquad\qquad-3\pd^Ig_{ab}g^{a\phi}w^bw^\phi+\frac32
\pd^Ig_{a\phi} g^{a\phi}w^\phi w^\phi\nn\\
&&\qquad\qquad-\frac32\varrho g_{ab}w^aw^\phi w^\phi
\pd^Iw^b+\frac12\varrho g_{a\phi}w^\phi w^\phi w^\phi
\pd^Iw^a\Big\}\,, \label{current.j2p.new}\\
J_{-2}^I&=&\frac{\sqrt{H g/\varrho}}{w^\phi w^\phi w^\phi
w^\phi}\Big\{-J_\varrho^I-w^\phi\pd^Iw^\phi +\pd^I
\ln\sqrt\varrho\;w^\phi w^\phi\nn\\
&&\qquad\qquad\quad-\pd^Ig_{ab}g^{a\phi}w^bw^\phi+\frac32
\pd^Ig_{a\phi} g^{a\phi}w^\phi w^\phi\nn\\
&&\qquad\qquad\quad+\frac12\varrho g_{ab}w^aw^\phi w^\phi
\pd^Iw^b-\frac32\varrho g_{a\phi}w^\phi w^\phi w^\phi
\pd^Iw^a\Big\}\,,\label{current.j2n.new}\\
J_\varrho^I&=&\frac3\varrho\Big(g^{\phi\phi} g_{a\phi}
\pd^Ig^{a\phi}-\pd^Ig^{\phi\phi}\Big)\,.\eea
%%%
The first two terms in both (\ref{current.j2p.new}) and
(\ref{current.j2n.new}) are of the sub-leading order and vanish on
the black hole horizons, but the contributions from the
$w^\phi\pd^Iw^\phi$ terms cancel the $\Delta'$ terms in
(\ref{divergence.j2}) exactly, while the contribution from
$J_\varrho^I$ is still negligible at the leading order.

So with (\ref{delta2p.new}) and (\ref{delta2n.new}), we have
$\hat\delta_{\pm2}$ acting as exact symmetries of the action
(\ref{action.BH}) on the black hole horizons. By using
(\ref{witt.generating}), we can obtain an infinite dimensional
conformal symmetry obeying the Witt algebra (\ref{algebra.witt}).
For each azimuthal angle $\phi^a$ with a non-vanishing angular
velocity, we will have an independent copy of the Witt algebra. So
classically, the action (\ref{action.BH}) has $k$-copies of
infinite dimensional conformal symmetries on the black hole
horizon, with $k$ being the number of non-vanishing angular
velocities. Since for a given classical solution there is no
essential difference between the reduced action (\ref{action.BH})
and the original action (\ref{action}), the same conclusion holds
for the original action (\ref{action}).

Note the conformal symmetries are fully determined by the
structure of the action (\ref{action.BH}) and the properties of
the background $H_{IJ}$, but are independent of the values of
$g_{ab}$, $w^a$ and $\varrho$. (For $\varrho
=\frac{v^2}{f^2}\cdot\frac{f}\Delta$, it is the factor
$\frac{v^2}{f^2}$ that should be treated as an independent degrees
of freedom, because the factor $\frac{f}\Delta$ is fixed in the
background.) One may entertain with the idea of treating
(\ref{action.BH}) as a field theory of $g_{ab}$, $w^a$ and
$\varrho$ living in the fixed background $H_{IJ}$, with the black
hole being the classical solution. Further, one can ask if the
fluctuations of the fields $g_{ab}$, $w^a$ and $\varrho$ can fully
describe the microstates of the black hole. We shall leave these
to future works.

\section{Summary}

In this paper, we have carried out a Kaluza-Klein like reduction
of the Einstein-Hilbert action along the ignorable coordinates of
stationary and axisymmetric black holes.  The reduced action
enables us to study the classical equations of motion in a much
greater detail. In the case of pure gravity plus a cosmological
constant, this allows us to re-derive the first law of black hole
thermodynamics in a straightforward manor.

The reduced action has a global $SL(m,R)$ gauge symmetry, with $m$
being the number of ignorable coordinates. Related to each angular
momentum there is a particular $SL(2,R)$ subgroup. We show that
this $SL(2,R)$ can be extended to the full Witt algebra on the
black hole horizons. The extended transformations are exact
symmetries of the actions (\ref{action.BH}) on the horizon. For a
black hole with $k$ non-vanishing angular velocities, the action
(\ref{action.BH}) then has $k$-copies of infinite dimensional
conformal symmetries on the horizon.

Our key motivation of the present work was to search a way that
can help us identify the conformal symmetries of the putative 2D
CFT dual to a non-extremal black hole, as suggested by the studies
of hidden conformal symmetries of black holes \cite{cms10}.
However, so far we have not been able to abstract any physical
information from the conformal symmetries found in this work. One
may try to reinterpret the extended symmetries (\ref{delta2p.g2}),
(\ref{delta2n.g2}) and (\ref{witt.generating}) as approximate
diffeomorphisms of the original action (\ref{action}) near the
horizons, and then use the usual covariant phase space method
(see, e.g. \cite{carlip11b} for the latest) to see if the Witt
algebra (\ref{algebra.witt}) can be promoted to a Virasoro algebra
at the quantum level. This procedure is still under investigation.

\section*{Acknowledgement}

Part of this work is benefited from conversations with Maria
Rodriguez, Oscar Varela, Dan Xie and Ilarion Melnikov. The author
also thanks Prof. Hermann Nicolai for a reference. This work was
supported by the Alexander von Humboldt-Foundation.

%\newpage

%%%%%%%%%%%%%%%%%%%%%%%%%%%%%%%%%%%
\end{document}